\documentclass[prb,nofootinbib,twocolumn,superscriptaddress]{revtex4} 
\usepackage[dvipdfm]{graphicx} 
\usepackage{dcolumn}% Align table columns on decimal point
\usepackage{bm}% bold math 
\usepackage{threeparttable}
\usepackage{times}
\usepackage{mathptmx}
\usepackage{lscape}
\usepackage{natbib}
\usepackage{amsmath}
\usepackage{braket}

\def\degree{\kern-.2em\r{}\kern-.3em}

\begin{document}

%\preprint{APS/123-QED} 

\title{ Theoretical study on density of microscopic states in configuration space via Random Matrix }

\author{Koretaka Yuge, Kazuhito Takeuchi and Tetuya Kishimoto}
\affiliation{
Department of Materials Science and Engineering,  Kyoto University, Sakyo, Kyoto 606-8501, Japan\\
}%

\begin{abstract}
In classical systems, our recent theoretical study provides new insight into how spatial constraint on the system connects with macroscopic properties, which lead to universal representation of equilibrium macroscopic physical property and structure in disordered states. 
These important characteristics rely on the fact that statisitcal interdependence for density of microscopic states (DOMS) in configuration space appears numerically vanished at thermodynamic limit for a wide class of spatial constraints, while such behavior of the DOMS is not quantitatively well-understood so far. The present study theoretically address this problem based on the Random Matrix with Gaussian Orthogonal Ensemble, where corresponding statistical independence is mathematically guaranteed. Using the generalized Ising model, we confirm that   lower-order moment of density of eigenstates (DOE) of covariance matrix of DOMS shows asymptotic behavior to those for Random Matrix with increase of system size.  This result supports our developed theoretical approach, where equilibrium macroscopic property in disordered states can be decomposed into individual contribtion from each generalized coordinate with the sufficiently high number of constituents in the given system, leading to representing equilibrium macroscopic properties by a few special microscopic states.
\end{abstract}

%\pacs{81.90.+c \sep 61.05.-a \sep 05.20.Gg \sep 05.10.-a \sep 02.30.Zz }

\maketitle

\section{Introduction}
In classical systems, statistical thermodynamics tells that macroscopic physical property (or macroscopic structure) in equilibrium state can be typically obtained by canonical average of $\overline{C} = Z^{-1}\sum_{d}C^{\left(d\right)}\exp\left(-E^{\left(d\right)}/k_{\textrm{B}}T\right)$, 
where $Z$ denotes partition function, $T$ is temperature, $d$ means possible microscopic state on phase space, and $E^{\left(d\right)}$ and $C^{\left(d\right)}$ are energy and physical property (or structure) in state $d$, respectively. 
When $T$ goes to zero, summation is performed for the ground states having lowest energy. When $T$ increases, due mainly to entropy contribution, the system can go into disordered states (e.g., states above the critical temperature). 
In the disordered states, direct determination of $\overline{C}$ through the definition of canonical average is practically intractable, since number of possible microscopic states should astronomically increases with increase of system size. Thus, a variety of calculation techniques have been developed to effectively address $\overline{C}$. One of the most successful techniques is Monte Carlo (MC) simulation with Metropolis algorism,\cite{metro} which samples important microscopic states contributing to $\overline{C}$, and the MC simulation have been subsequently  modified such as multihistgram method, multicanonical ensembles and entropic sampling,\cite{mc1,mc2,mc3} in order to effectively sweep across the configuration space. 
One of the exceptions is coherent potential approximation,\cite{cpa} where it considers the average occupation of elements with the lack of information about geometrical structure. 
Another is high-temperature expansion,\cite{ht} enabling efficient estimation of energy as well as other physical properties at high temperature. 

When we consider classical systems, typically their constituents are spatially constrained in various ways. For instance, crystal lattice acts as constraint for substitutional crystalline solids, and volume and density as constraints for liquids in rigid box. Although the existing theoretical approaches amply predict macroscopic physical property and structure in equilibrium metallic and semiconductor alloys based on first-principles calculations,\cite{i5,i6,i7,i9,i10,i11,i13} the role of the spatial constraint on equilibrium properties do not get sufficient attention so far. Following this fact, our recent studies\cite{lsi,si} find new representation of macroscopic physical properties (including internal energy, density, and free energy) and of macroscopic structure for disordered states, by using a few number of specially selected microscopic states. We find that these special microscopic states depend only on the type of spatial constraint, and are independent of constituent elements, multibody interactions, and of temperature, which means that we can \textit{a priori} know the structure of the special microscopic states. We have demonstrated that this finding provides significantlly efficient and systematic prediction of equilibrium properties for multicomponent alloys based on first-principles calculation. 
This important characteristics for equilibrium properties rely on the fact that statisitcal interdependence for density of microscopic states (DOMS) in configuration space appears numerically vanished at thermodynamic limit for a wide class of spatial constraints, whereas such behavior of the DOMS have not been quantitatively well-understood so far. In the present article, we theoretically tuckle this problem based on the Random Matrix with Gaussian Orthogonal Ensemble, where corresponding statistical independence is mathematically guaranteed.

\section{Method and Discussions}
\begin{figure}[h]
\begin{center}
\includegraphics[width=0.88\linewidth]{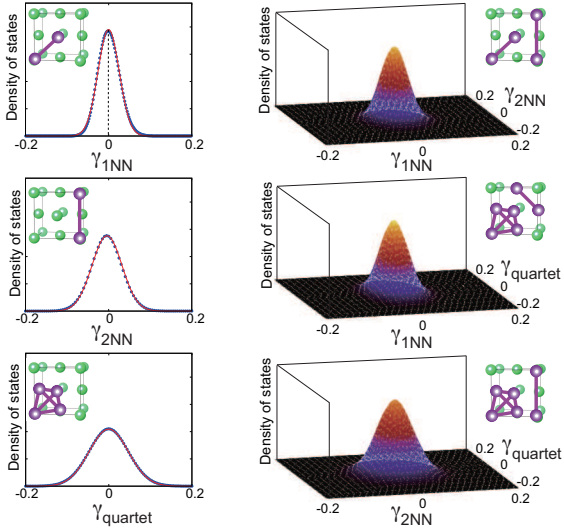}
\caption{Left: Density of microscopic states (DOMS) along single-variate coordination of nearest, second-nearest pairs and quartets for A$_{128}$B$_{128}$ binary system on fcc lattice. Right: DOMS for bivariate coordination composed of two of the three figures in left. Closed circles or dotted lines denote result from simulation uniformly sampling the configuration space, and solid curves denote analytical gaussian distributions. }
\label{fig:dos}
\end{center}
\end{figure}

  Let us first briefly explain the density of microscopic states (DOMS) on configuration space for \textit{non-interacting} system. As shown in Fig.~\ref{fig:dos}, we find in the present study that for a wide class of spatial constraints, DOMS for large number of constituents with multicomponent system can be universally well-characterized by a multidimensional gaussian distribution, namely,
\begin{eqnarray}
g\left(q_{1},\ldots,q_{g}\right) \simeq \frac{1}{2\pi^{\frac{g}{2}} \left|\pmb{\Gamma}\right| ^{1/2}}\exp\left[-\frac{1}{2}\cdot\pmb{H} \pmb{\Gamma} ^{-1} \pmb{H}^{\textrm{T}}  \right], 
\end{eqnarray}
where $\left\{q_{1},\ldots,q_{g}\right\}$ denotes complete basis functions to describe possible microscopic states, $\pmb{H}$ is the 
$g$-component vector of $\left(q_{1}-\Braket{q_{1}}_{1},\ldots,q_{g}-\Braket{q_{g}}_{1}\right)$ where $\Braket{\quad}_{1}$ denotes taking arithmetic average over configuration space, and $\pmb{\Gamma}$ represents covariance matrix for $g\left(q_{1},\ldots,q_{g}\right)$, reflecting the spatial constraint. 
  This means that under a proper set of basis functions to give diagonal $\pmb{\Gamma}$, statistical interdependence practically vanishes, although in principle, basis functions themselves are not statistically independent due to the existence of spatial constraint.\cite{ss} 
  
  In order to address this practical disappearance of statistical interdependence, straightforward comparison of DOMS landspace for real systems and that for analytical gaussian distribution is practically intractable: This is because deviation in the landscapes is negligiblly small, which can be easily buried by numerical error in simulation to sweep across the vast configuration space. 
  In the present study, to avoid this problem, we therefore take different strategy to employ Random Matrix with Gaussian Orthogonal Ensemble, whose statistical independence is mathematically guaranteed. The random matrix, $\mathbf{R}$, is given by $l\times m$ matrix whose elements independently take normal random numbers, with its linear average and variance respectively taking 0 and 1. In order to apply the random matrix to investigating characteristics of DOMS, we read $l$ as number of uniformly sampling point on configuration space for real system, and $m$ as number of basis functions considered. With this definition, covariance matrix $\pmb{\Gamma}_{\textrm{R}}$ of DOMS constructed from the random matrix is naturally given by
\begin{eqnarray}
\pmb{\Gamma}_{\textrm{R}} = \frac{1}{l}\mathbf{R}^{T}\mathbf{R},
\end{eqnarray}
where all diagonal element is normalized to 1. 

  For real system, we employ generalized Ising model\cite{ce} on a given lattice, where any microscopic state on configuration space, $\left|\phi\right>$, can be completely specified by 
\begin{eqnarray}
\left|\phi\right> = \sum_{p}\left|p\right>\Braket{p|\phi}.
\end{eqnarray}
In binary system where occupation of elements is specified by Ising-like spin of $\sigma\pm 1$, coefficient of the basis function $\left|p\right>$ can be simply given by
\begin{eqnarray}
\Braket{p|\phi} = \Braket{\prod_{k\in \alpha_{p}}\sigma_{k}}_{\phi},
\end{eqnarray}
where product $k$ is taken for ``figure'' $\alpha_{p}$ along $\left|p\right>$, consisting of corresponding lattice points, and 
$\Braket{\quad}_{\phi}$ denotes linear average over symmetry-equivalent figure to $\alpha_{p}$ on microscopic state $\left|\phi\right>$.
  The advantage to employ the generalized Ising model is that for binary system, we confirm that the basis functions above can provide diagonal $\Gamma$ for any combination of corresponding figures at equiatomic composition: We do not need to further find a set of basis functions to deagonalize $\Gamma$. 
  Here, we therefore prepare equiatomic binary system on fcc lattice, where basis functions along up to 6th nearest-neighbor (6NN) pairs and all triplets consisting of up to 6NN pairs are considered, resulting in 29 basis functions. 
In a similar fashion to the random matrix, we construct DOMS constructed from the 29 basis functions by performing Monte Carlo (MC) simulation to uniformly sampling microscopic states (500,000 MC step) for 256, 2048, and 3024-atom MC-cell to see the system-size dependence of the DOMS. 
Variance of DOMS projected onto individual coordination is all normalized to one so that diagonal elements of resultant covariance matrix all take one. 

\begin{figure}[h]
\begin{center}
\includegraphics[width=0.90\linewidth]{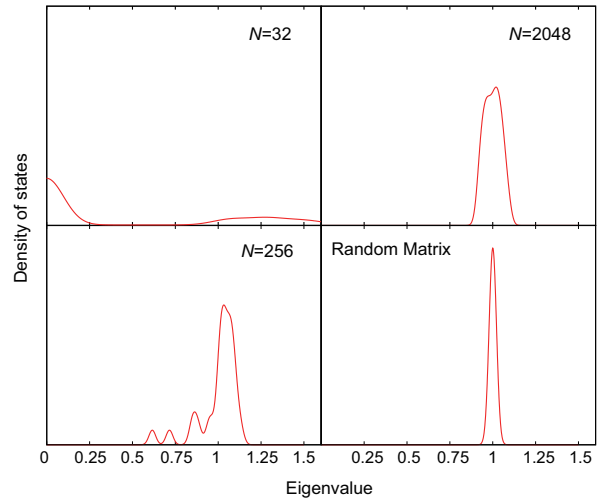}
\caption{Density of states for eigenvalues of covariance matrix, constructed from random matrix and from practical system (equiatomic composition on fcc lattice) with number of constituents in the system, $N$, takes 256, 2048 and 3200.   }
\label{fig:doe}
\end{center}
\end{figure}

  As is already described above, since direct comparison of landscape of DOMS is intractable, we compare the density of eigenvalues (DOE) of the covariance matrix obtained through random matrix and practical system. 
  For random matrix, we first numerically construct $l\times m$ random matrix where constituent elements are obtained by normal numbers as described above. Here, to make comparison with the practical system, we respectively give $l=29$ and $m=500,000$.  To check the validity, we compare the DOEs of the constructed Random Matrix and those obtained through the characteristics of Marcenko-Pastur distribution, where corresponding DOE of covariance matrix can be analytically determined with $l\to\infty$  and $m\to\infty$, where the ratio of $l/m$ is kept fixed. We confirm that the DOEs for the constructed Random Matrix discussed below exhibits excellent agreement with the analytical ones.

\begin{figure}[h]
\begin{center}
\includegraphics[width=0.76\linewidth]{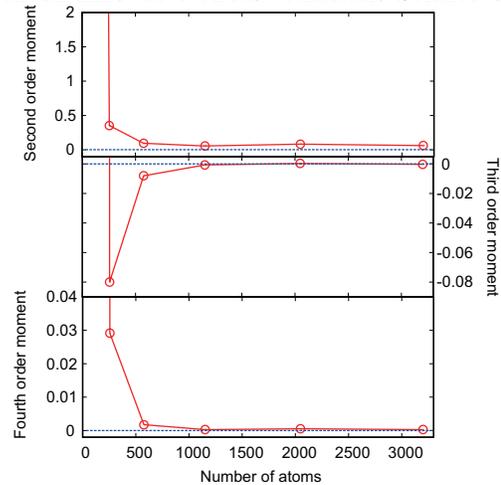}
\caption{2nd, 3rd and 4th-order moment of DOEs for practical system as a function of number of constituents. The dashed lines denote corresponding moments of DOEs obtained through random matrix. }
\label{fig:mom}
\end{center}
\end{figure}

  Figure~\ref{fig:doe} compares the resultant DOEs for random matrix and for practical systems with different number of constituents, $N$. 
 We can see that when the system size is small (i.e., $N=32$), landscape of the DOE appears completely different from that for Random Matrix. Especially, density of states around $\epsilon=0$  comes from the significant statistical interdependence between chosen basis functions due to the periodic boundary condition.
  Meanwhile,  when $N$ increases, landscape of DOEs for practical systems appears gradually close to that for random matrix having single sharp peak with non-zero finite width around $\epsilon=1$. Particularly, several sub-peaks below $\epsilon=1$ for practical systems are significantly diminished when $N$ increases. The highest peaks of DOEs around $\epsilon=1$ for random matrix as well as for practical systems can be naturally attributed to the fact that trace of covariance matrix for random matrix and for practical systems take exactly the same value. 
  To further make quantitative comparison, we estimate lower-order (from 2nd to 4th) moment of the DOEs.  Figure~\ref{fig:mom} shows system-size dependence of the resultant 2nd, 3rd and 4th moment of DOEs for practical system, together with that for random matrix represented by dashed horizontal lines. We can clearly see that when $N$ increases, value of all moments for practical system approaches to that for random matrix. 
  These results therefore indicate that when the number of microscopic states around the center of gravity of DOMS in configuration space increase with increase of the system size, statistical interdependence along each coordination can be vanished: This certainly supports our developed approach using a few special microscopic states, which uses this characteristics of statistical interdependence to decompose contribution from each coordinate to macroscopic physical properties.

To summarize, by using the characteristics of random matrix consisting of independently random numbers, we confirm that statistical interdependence of density of microscopic states in configuration space tend to disappear when the system size increases. This result support 
our developed approach to provide new representation for equilibrium macroscopic properties, based on the dissapearance of the statistical interdependence for large systems.

\section*{Acknowledgement}
This work was supported by a Grant-in-Aid for Scientific Research on Innovative Areas 
``Materials Science on Synchronized LPSO Structure'' (26109710) and a Grant-in-Aid for Young Scientists B (25820323) from the MEXT of Japan, Research Grant from Hitachi Metals$\cdot$Materials Science Foundation, and Advanced Low Carbon Technology Research and Development Program of the Japan Science and Technology Agency (JST).

\end{document}